\documentclass[12pt, preprint]{elsarticle}

\usepackage{hyperref}

\journal{Combustion and Flame}

\biboptions{sort&compress}
\usepackage{amsmath,amssymb}
\usepackage{color}
\usepackage{mathtools, cuted}

\usepackage[margin=1in]{geometry}

\newcommand{\Gr}{\text{\textit{Gr}}}
\newcommand{\Ra}{\text{\textit{Ra}}}

\newcommand{\ie}{i.e.\ }

\newcommand{\ms}{\kern.10em\relax}

\newcommand{\cm}{\,\text{cm}}
\newcommand{\mm}{\,\text{mm}}

\definecolor{cardinal}{rgb}{0.77, 0.12, 0.23}

\usepackage{blindtext}

\usepackage{layouts}
\begin{document}

\begin{frontmatter}

\title{Observed dependence of characteristics of liquid-pool fires on swirl magnitude}
\tnotetext[mytitlenote]{}

\author[ucsd]{W.~Coenen}
\cortext[mycorrespondingauthor]{Corresponding author}
\ead{wicoenen@ucsd.edu}

\author[ucsd]{E.~J.~Kolb}
\author[ucsd]{A.~L.~S\'anchez}
\author[ucsd]{F.~A.~Williams}

\address[ucsd]{Department of Mechanical and Aerospace Engineering, %
               University of California San Diego, %
               La Jolla, CA 92093-0411, USA}

\begin{abstract}
One dozen vertically oriented thin rectangular vanes, 62 cm tall and 15.2 cm wide, were placed
27 cm from the center of heptane and ethanol pool fires in continuously fed, floor-flush pans
3.2 cm and 5.1 cm in diameter in the laboratory. The vanes were all oriented at the same fixed
angles from the radial direction, for 9 different angles, ranging from 0 to 85 degrees, thereby
imparting 9 different levels of circulation to the air entrained by each pool fire. The
different swirl levels were observed to engender dramatically different pool-fire structures.
Moderate swirl suppresses the global puffing instability, replacing it by a global helical
instability that generates a tall fire whirl, the height of which increases with increasing
circulation. Except for the largest heptane pool, higher swirl levels produced vortex
breakdown, resulting in the emergence of a bubble-like recirculation region with a ring vortex
encircling the axis. Measured burning rates increase with increasing swirl levels as a
consequence of the associated increasing inflow velocities reducing the thickness of the
boundary layer within which combustion occurs right above the liquid surface, eventually forming
detached edge flames in the boundary layer that move closer to the axis as the circulation is
increased. Still higher circulation reduces the burning rate by decreasing the surface area of
the liquid covered by the flame, thereby reducing the height of the fire whirl. Even higher
circulation causes edge-flame detachment, resulting in formation of the blue whirl identified in
recent literature, often meandering over the surface of the liquid in the present experiments.
This sequence of events is documented herein.
\end{abstract}


\end{frontmatter}


\section{Introduction}
\label{sec:introduction}

Aside from the inherent thermochemical and transport parameters associated with the fuel, the
gas-phase motion induced by buoyancy, for an axisymmetric pool fire of a liquid contained in a
round pool of radius $a$, is controlled by a single dimensionless number, the Rayleigh number
$\Ra=(g a^3)/(\nu D_T)$, defined here with use of the kinematic viscosity $\nu$ and thermal
diffusivity $D_T$ of the ambient air, with $g$ denoting the gravitational acceleration;
alternatively, the associated Grashof number $\Gr=(g a^3)/\nu^2$ could have been selected,
since the Prandtl number $\nu/D_T$ is practically constant.
Imposition of circulation of air about the axis introduces a second controlling parameter, which
may be taken to be the inlet angle $\alpha$ between the inward radial component of velocity and
the circumferential component of velocity at a specified, geometrically similar, external
location in the surrounding air. The purpose of the present investigation is to explore the
influence of this second controlling parameter on the structure and dynamics of axisymmetric
pool fires, thereby helping to improve our understanding of fire whirls, which have been
thoroughly reviewed recently~\cite{Tohidi.etal.2018}.

A distinctive characteristic of liquid-pool flames is their tendency to develop self-sustained
oscillations, shedding large toroidal coherent structures at a well-defined frequency while
maintaining axial symmetry, a phenomenon referred to in the literature as ``puffing''. It is
well established that, under normal conditions of temperature and pressure, diffusion flames
formed over hydrocarbon fuel pools larger than a few centimeters in diameter puff with a
frequency on the order of $f \sim \sqrt{g/a}$ \cite{Emmons.1980, Heskestad.1998, Joulain.1998,
Tieszen.2001}. Despite earlier opinions to the contrary, favoring a convective
instability~\cite{Tieszen.2001}, pool-fire puffing recently has been definitively
shown~\cite{MorenoBoza.etal.2018} to correspond to a hydrodynamic global instability of the
flow, the onset of which is associated with a critical value of the Rayleigh number.

The puffing phenomenon of pool fires is not seen in fire whirls, which exhibit quite different
types of fluid flow~\cite{Tohidi.etal.2018}, more closely resembling then flow in fully
developed turbulent buoyant plumes. The entrainment rate of the turbulent plume developing above
the flame, linearly proportional to the burning rate, increases with the two thirds power of the
vertical distance, inducing in the surrounding atmosphere a flow that is largely inviscid, with
the exception of a thin boundary-layer region adjacent to the horizontal
wall~\cite{Turner.1969}. The deflection of the induced radial inflow at geometrical constraints
found in the far field, such as vertical vanes in experiments or topographic features in
wildland fires, may introduce an azimuthal velocity component, leading to the establishment of
swirling motion. The centripetal acceleration of the resulting swirling flow, whose magnitude
$\Gamma^2/r^3$ is proportional to the square of the flow circulation $\Gamma$ and inversely
proportional to the cube of the distance to the axis $r$, is balanced by the radial pressure
gradient. Viscous forces slow down the swirling motion in a near-wall boundary layer, where the
imposed pressure gradient generates a strong radial inflow with characteristic velocities $v_r$
increasing on approaching the axis according to $v_r \sim \Gamma_o/r$, where $\Gamma_o$ is the
value of the circulation outside the boundary layer~\cite{Burggraf.etal.1971}.
As suggested by previous experimental results \cite{Dobashi.etal.2016}, the collision of
this wall jet over the fuel pool plays a key role in the formation of the columnar vortex at the
base of the fire whirl.

A number of different approaches have been employed to generate fire whirls in the laboratory
for visual observations and scientific measurements. The earliest design employed two slightly
offset, vertically oriented, semi-cylindrical surfaces surrounding the pool that leave two small
slits for the tangential inflow of the entrained air~\cite{Byram.Martin.1962}. This
configuration, which has often been used for scientific measurements, extending to recent
times~\cite{Hartl.Smits.2016, Kuwana.etal.2011}, or variations of it involving offset vertical
planar walls, can be quite convenient but does not afford easy access to investigating effects
of different inlet angles $\alpha$. That angle can be easily adjusted if the circular walls are
replaced by a rotating cylindrical screen~\cite{Emmons.Ying.1967}, but that requires
construction of a more elaborate, motor-driven apparatus. It is more convenient to employ a set
of long vertical flow vanes placed at an adjustable angle with respect to the radial direction,
as was done in an experimental study of dust devils~\cite{Mullen.Maxworthy.1977}. This last
approach is the one selected here. As a consequence of this selection, the results reported
below, including the specific sequence of flow phenomena occurring for increasing $\alpha$,
pertain to fire whirls where the swirling motion is driven by plume entrainment, as often occurs
in realistic scenarios, and might be different from those observed in experimental
configurations involving rotating screens or tangential blowers, for example, for which the
level of swirl attainable is independent of the near-axis flow entrainment.

Just as for any other slender swirling flow, dust devils and fire whirls may be subject to
vortex breakdown \cite{Hall.1972, Escudier.1988} (an abrupt change in the flow structure with a
very pronounced retardation of the streamwise flow leading to the formation of a stagnation
point along the axis and a corresponding divergence of the stream surfaces). Recent experiments
\cite{Xiao.etal.2016} have unveiled the existence of a new type of whirling flame, developing
after vortex breakdown, with a fundamentally different structure consisting of a light blue
inverted cone at the base, extending outward to a bright ring, followed by a purple haze above.
This so-called \textit{blue whirl} has been found as a transient object evolving naturally from
a traditional fire whirl as the circulation is increased \cite{Hariharan.etal.2019a}. Its
existence has been shown to rely on the presence of the recirculating-flow bubble that results
from the vortex breakdown, which can develop only for sufficiently large ambient circulation
\cite{Hu.etal.2019}. All of these phenomena are encountered in the present study.


\section{The experimental arrangement}
\label{sec:experiment}

\begin{figure}[tb]
    \centering
    \includegraphics[width=252pt]{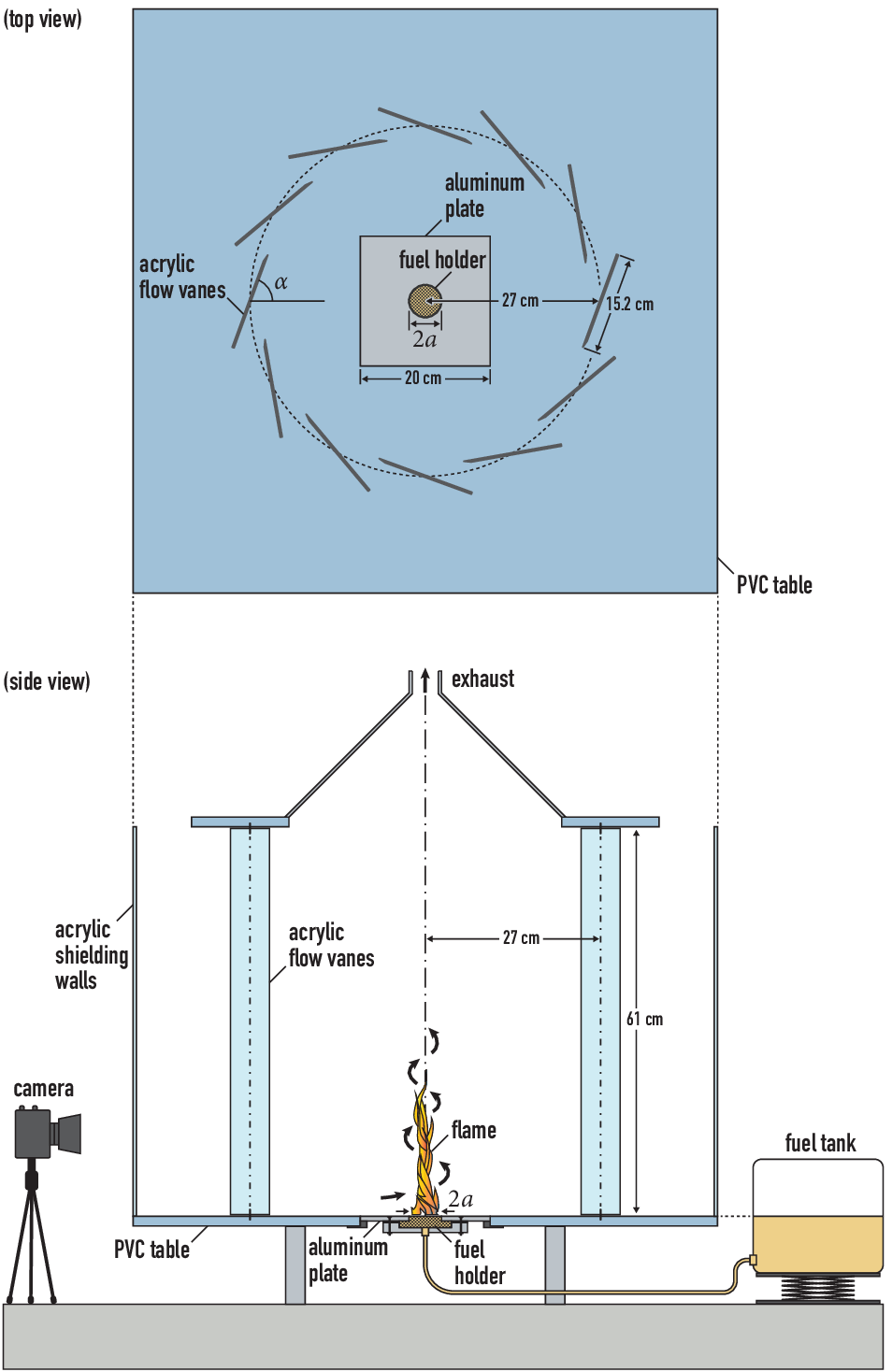}
    \caption{%
        Schematic view of the experimental setup used in this study.}
    \label{fig:expsetup}
\end{figure}

In our experiments, the inclination angle $\alpha$ of the vanes with respect to the radial
direction (see Fig.~\ref{fig:expsetup}) is used as a direct controlling mechanism to set the
level of ambient circulation, with $\Gamma \propto \tan \alpha$. The case $\alpha = 0$
corresponds to a swirl-free atmosphere, and it results in puffing flames whose stability
characteristics have been studied recently~\cite{MorenoBoza.etal.2018}. For $\alpha$ close to
$\alpha = 90^\circ$, the resulting outer flow conditions can be expected to approach those
present in experiments using two slightly offset semi-cylindrical surfaces. The entire range of
swirl levels and associated stationary swirling-flame solutions are to be explored here by
sequentially modifying the value of $\alpha$ in a series of experiments for a given fuel and a
given pool radius.

The experimental facility, depicted in Fig.~\ref{fig:expsetup}, was designed using as a basis
the setup recently employed to study the onset of pool-fire puffing~\cite{MorenoBoza.etal.2018}.
A set of twelve vertical acrylic vanes with a width of $15.2\cm$ and a height of $61\cm$ was
placed at a large radial distance to deflect the radial air inflow induced by the entrainment of
the flame and induce a swirling motion. Although this method has been employed to produce swirl
in seminal experiments of dust-devil dynamics~\cite{Mullen.Maxworthy.1977}, it does not appear
to have been employed in fire-whirl experiments. The angle of inclination of the vanes with
respect to the radial direction, $\alpha$, serves to control the level of swirl, as previously
discussed. The burner, which sits in the center of a $90\cm\times90\cm\times12.7\mm$ PVC table,
is composed of a $20\cm\times20\cm\times6.35\mm$ aluminum plate with a circular hole of diameter
$2a = 32\mm$ or $2a = 51\mm$ in its center, and an aluminum fuel holder mounted underneath.

To prevent convective currents in the liquid fuel, the fuel holder is filled nearly to the rim
with a layer of glass beads ($3{\rm mm}$ in diameter). The fuel holder is connected to an
external fuel tank with a cross-sectional area of approximately $350\,\cm^2 \gg \pi a^2$. In
this manner, once the height of the tank is adjusted so that the burner is filled to the desired
level with fuel, this level remains essentially constant during the course of a measurement
(typically about five minutes). The external tank is placed on a high-precision load cell to
allow measurement of the fuel-consumption rate.

External perturbations to the flame are minimized by surrounding the complete experiment
with acrylic shielding walls, the volume between these walls and the vanes, as well as the
opening at the top, being large enough that the air contained therein differs negligibly from
that which would be present in a quiescent ambient atmosphere. The flame dynamics corresponding
to a given swirl level was recorded with a Panasonic Lumix FZ300 camera at a frame rate of
$120\,\text{fps}$. At each angle $\alpha$, the flame is allowed to burn for a few minutes so
that the system reaches a steady state before beginning simultaneous data acquisition of flame
behavior and burning rate. The latter was obtained by averaging the fuel consumption over
one-minute intervals, with differences in burning rates measured over consecutive intervals
remaining always smaller than 5\%, hence confirming the steady operation of the system.


\section{Experimental results and discussion}
\label{sec:results}

\subsection{Sample flame images and associated burning rate}

\begin{figure*}
    \centering
    \includegraphics[width=522pt]{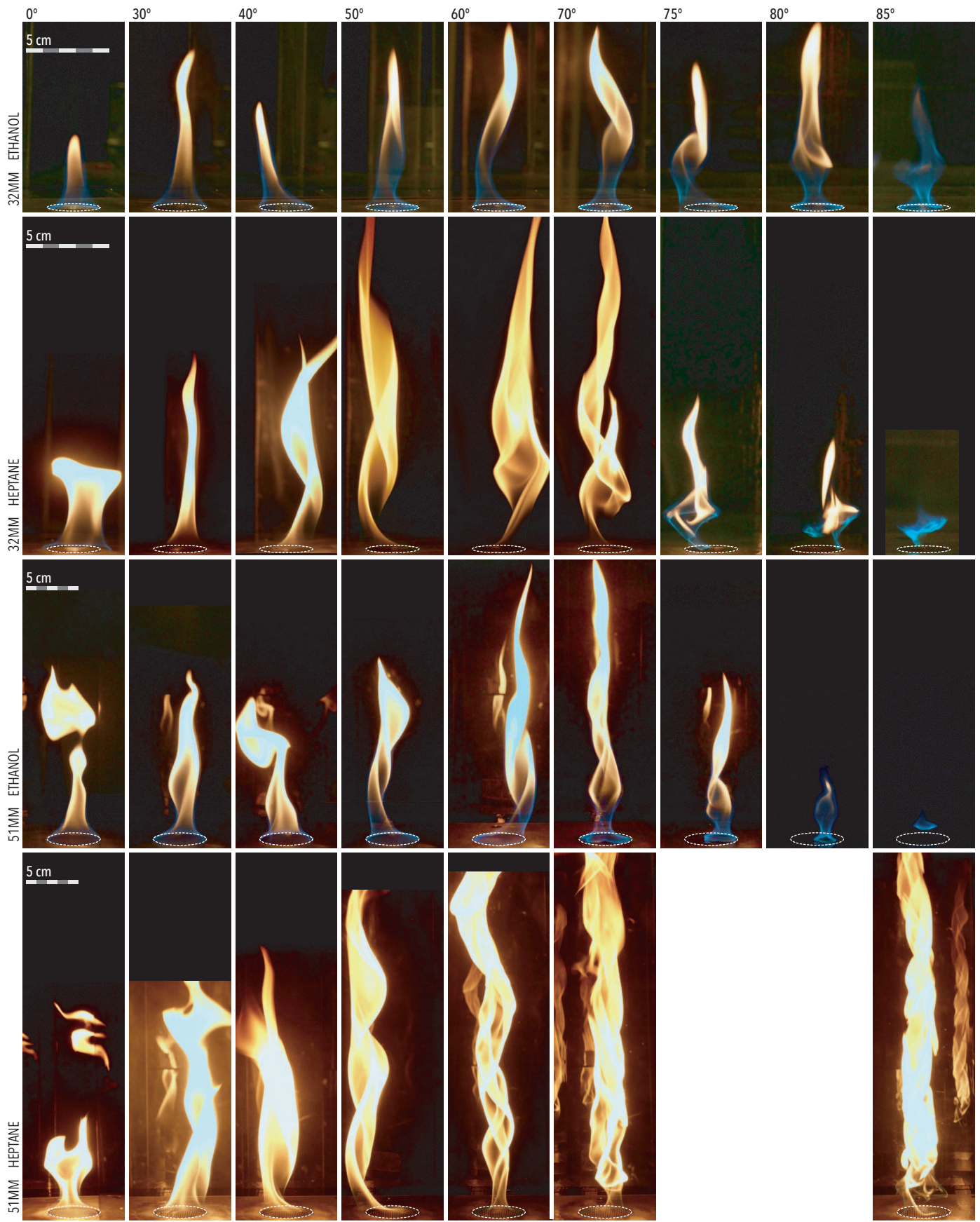}
    \caption{%
        Sample flame images for two fuels, two fuel-pool diameters, and a range of
        flow-vane inclination angles.}
    \label{fig:shapes}
\end{figure*}

Results were obtained for two fuel-pool diameters, $2a = 32\mm$ ($\Ra = 111000$) and $2a =
51\mm$ ($\Ra = 451000$), and values of the flow-vane inclination angle in the range  from
$0^\circ$ to $85^\circ$. To test effects of radiant energy flux, experiments were conducted for
two fuels with different sooting propensity, namely ethanol (low propensity) and heptane (high
propensity). For the pool sizes considered, the resulting flames are always unsteady.  A summary
of sample flame instantaneous images, selected from the videos to illustrate the different flow
phenomena occurring at each inclination angle, is presented in Fig.~\ref{fig:shapes}. As the
inclination angle is increased, the flames are seen in this figure to experience transitions
through a series of stages, each of which is described in the following sub-sections.

\begin{figure}[tb]
    \centering
    \includegraphics[width=252pt]{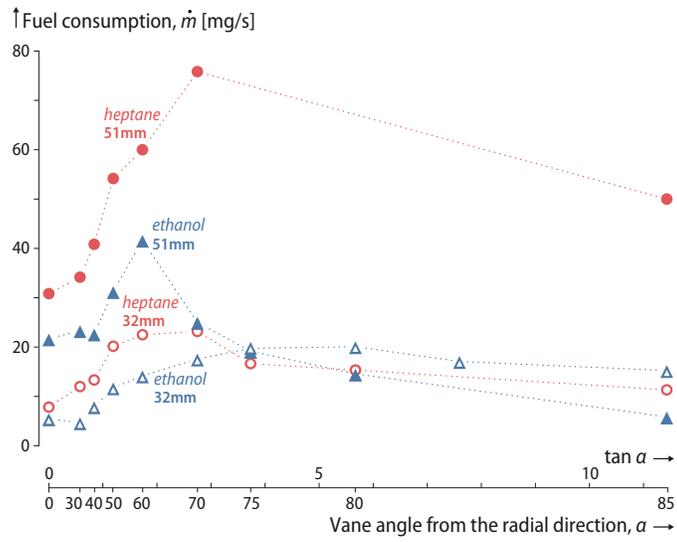}
    \caption{%
        Fuel consumption as a function of ambient swirl, for heptane and ethanol, and two
        different fuel-pool diameters. The experimental measurements, represented by symbols,
        are connected by dashed lines to facilitate visualization of the results.}
        \label{fig:fuelcons}
\end{figure}

Fig.~\ref{fig:fuelcons} shows the measured fuel consumption rates associated with the
experiments in Fig.~\ref{fig:shapes}. With increasing levels of ambient swirl (\ie increasing
values of the inclination angle $\alpha$), the fuel consumption first increases, then reaches a
maximum, beyond which it is seen to decrease again. The initial increase is associated with an
increased rate of heat transfer to the fuel surface caused by the flame sheet lying closer to
the fuel surface. The flame, attached to the rim of the pan, is in a boundary layer that becomes
thinner as $\alpha$ increases. The final decrease in fuel-consumption rate is associated with a
detachment of the flame edge from the rim, effectively reducing the rate of fuel vaporization
(see section 3.4). The higher burning rate of heptane than ethanol results from its higher
heat-transfer rate.

\subsection{Helical global modes}

For both values of the fuel-pool diameters used in this set of experiments the associated
Rayleigh numbers are supercritical in the absence of ambient circulation (\ie for $\alpha = 0$),
so that the resulting flame puffs in a nearly axisymmetric fashion, as predicted by the global
instability analysis~\cite{MorenoBoza.etal.2018}. The global puffing instability is suppressed
by the appearance of a stationary helical global instability as the level of swirl in the
surrounding atmosphere increases for increasing $\alpha$. A strong puffing-free fire whirl is
eventually seen to form, around $\alpha \simeq 50^\circ$.

These observations indicate that the formation of a fire whirl requires a sufficient level of
ambient swirl, corresponding to a sufficiently large value of the inclination angle $\alpha$.
This is consistent with the present knowledge of liquid-pool-fire stability. Thus, in the
absence of ambient swirl, the global hydrodynamic instability of liquid-pool fires is dominated
by the azimuthal mode number $m=0$ (axisymmetric), the puffing mode \cite{MorenoBoza.etal.2018},
resulting in the puffing flame shown for $\alpha=0$ in Fig.~\ref{fig:shapes}. Under these
swirl-free conditions, values of $m \ge 1$ (helical) modes display lower growth rates and
consequently play a secondary role in the resulting flame dynamics. The axisymmetric instability
mode is also dominant in jet diffusion flames~\cite{MorenoBoza.etal.2016}, low-density
jets~\cite{Coenen.etal.2017}, and buoyant
plumes~\cite{Bharadwaj.Das.2017,Chakravarthy.etal.2018}.

The presence of ambient swirl may change this stability behavior. It may be expected that the
growth rate of the axisymmetric ($m=0$) and helical ($m \ge 1$) modes carries a dependence on
the level of ambient swirl, additional to the dependence on the Rayleigh number $\Ra$. The
transition from puffing to whirling therefore is likely to be associated with the critical swirl
level at which the growth rate of the helical mode (possibly $m=1$) exceeds that of the
axisymmetric mode. Global stability analyses accounting for the presence of ambient swirl, which
are not yet available, would be needed to test this inference, providing accurate quantification
of the transition conditions. It is known that such modes exist, but it has not been established
that any of them can reach instability at lower Rayleigh numbers than that for the axial mode.

The higher ambient circulation associated with larger values of $\alpha$ is accompanied by a
larger radial pressure gradient, which, in turn, accelerates radially the gas in the near-wall
boundary-layer region surrounding the base of the fire. This affects the diffusion flame that
develops from the pool rim, causing it to approach the fuel-pool surface, increasing the
associated fuel vaporization rate and resulting in stronger fire whirls with larger heights.

\subsection{Vortex breakdown}

\begin{figure}[tb]
    \centering
    \includegraphics[width=252pt]{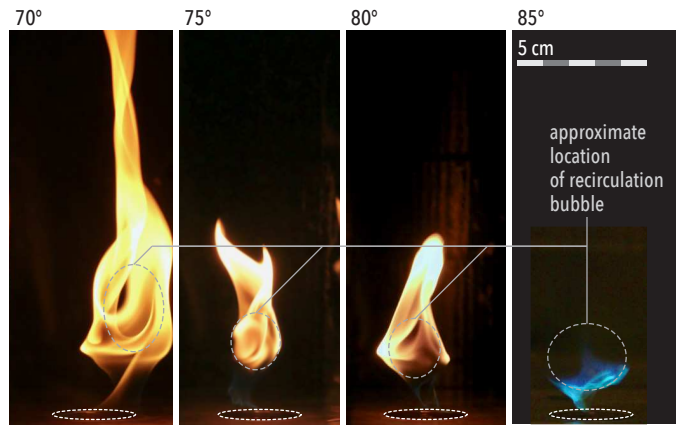}
    \caption{%
        The appearance of a recirculation bubble as a consequence of vortex breakdown, for
        heptane with a fuel-pool diameter $2a = 32\mm$.}
    \label{fig:vortexbd}
\end{figure}

An additional increase of the inclination to $\alpha=70^{\circ}$ further lengthens the fire
whirl, with the resulting strong swirling motion leading to vortex breakdown of the columnar
flame, indicated by the emergence of a bubble-like recirculating region (see
Fig.~\ref{fig:vortexbd}). Although vortex breakdown has been known to be present in fire whirls
\cite{Emmons.Ying.1967} and dust devils \cite{Mullen.Maxworthy.1977}, the specific conditions
needed for its development and the resulting effects on the flow are still not fully understood.
The associated problem, however, is fundamentally different from that of vortex breakdown in
technological applications pertaining to combustion, in which the swirling motion is imparted in
the jet stream \cite{Candel.etal.2014}, with the ratio of the fluxes of angular momentum to
axial momentum defining a swirl number that characterizes vortex breakdown~\cite{Hall.1972}.
A critical swirl number of order unity defines in these combustors the transition from a slender
jet to a vortex-breakdown flow with a stagnation point along the axis.

In contrast to swirl combustors, the definition of a swirl number determining the critical
conditions for formation of vortex breakdown in fire whirls is not clear. For fire whirls, the
breakdown bubble forms in the near-pool region where the incoming boundary-layer flows at all
angles collide at the center to give an uprising jet flow \cite{Mullen.Maxworthy.1977}. The
presence of a stagnation point deflecting the flow reduces the axial momentum flux, thereby
limiting the resulting fire-whirl height, as shown in Fig.~\ref{fig:vortexbd}. Present
experimental observations appear to indicate that the bubble moves towards the fuel-pool surface
as the value of $\alpha$ is increased further. The conditions for vortex breakdown to occur in
fire whirls, as well as the subsequent history of the breakdown bubble, are in need of further
investigation. Such studies may clarify why breakdown did not occur for the larger heptane fire,
which exhibited a substantially higher burning rate and radiant energy flux level, although
breakdown may have been seen if it had been possible to place the vanes at a larger distance
from the axis.

\subsection{Fire-whirl height reduction}

The anchoring of the flame edge at the pool rim depends critically on the strain rate imposed by
the incoming radial flow, with higher strain rates for increasing values of $\alpha$. As
explained in the introduction, the results of an early theoretical analysis pertaining to the
boundary-layer flow induced by a potential vortex \cite{Burggraf.etal.1971} suggest that the
flow surrounding the fire whirl includes a near-wall radial jet with characteristic radial
velocity $v_r \sim \Gamma_o/r$; the associated near-wall velocity gradients at first decrease
with decreasing radius through boundary-layer growth, then reach a minimum, and finally increase
according to $\sim \Gamma_o/r^2$ as the radius approaches zero, where $\Gamma_o$ is the value of
the circulation at the outer edge of the boundary layer. As can be expected from edge-flame
theory~\cite{Linan.etal.2015}, the flame may remain attached provided that the strain rate at
the rim of the fuel pool, of order $\Gamma_o/a^2$, does not exceed a critical value, given in
order of magnitude by the inverse of the residence time across the stoichiometric planar
deflagration~\cite{Linan.etal.2015}. Since in the experiments $\Gamma_o \propto \tan \alpha$, a
sufficiently large value of $\alpha$ should cause detachment of the edge flame from its near-rim
anchoring region. The location of the flame edge over the fuel surface after detachment becomes
determined by a balance between the edge flame-spread propagation velocity
\cite{Hirano.Saito.1991,Ross.1994} and the boundary-layer radial inflow.

As can be seen in Fig.~\ref{fig:shapes}, all flames with $\alpha \le 60^\circ$ remain attached.
For larger values of $\alpha$ the edge flame recedes from the pool edge, moving radially inward
to a location over the fuel-pool surface where it encounters a lower strain rate, as seen for
$\alpha=70^\circ$ and, more clearly, for $\alpha=75^\circ$. The reduced extent of the flame base
results in a smaller global fuel-vapor vaporization rate, significantly limiting the height of
the associated flame, as seen most dramatically in the flow established for $\alpha=80^\circ$.
The resulting shortened fire whirl moves continuously around the axis as the edge-flame wanders
in response to the boundary-layer flow in these experiments. Although differently designed
experiments might eliminate this wandering, there is nothing in the present tests that could
stabilize whatever in-plane instabilities promote the observed meander.

\begin{figure*}[tb]
    \centering
    \includegraphics[width=522pt]{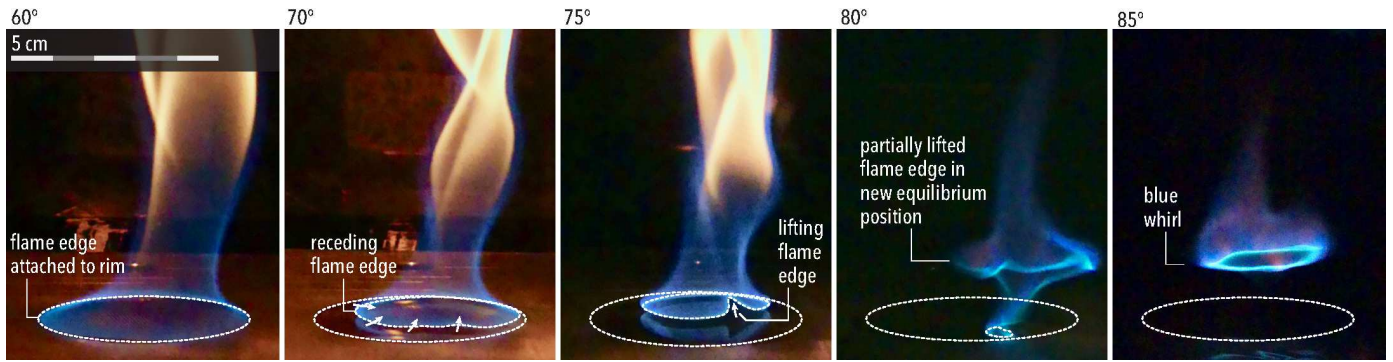}
    \caption{%
        Detachment of the flame edge from the rim, lifting of the flame edge, and the
        formation of a blue whirl, for ethanol with a fuel-pool diameter $2a = 51\mm$.}
    \label{fig:edge}
\end{figure*}

Shown in Fig.~\ref{fig:edge} are close-up views at the base of the larger of the two ethanol
fires, for conditions corresponding to the last five angles of Fig.~\ref{fig:shapes}. In the
first frame at the left, with $\alpha \simeq 60^\circ$, the flame is seen to remain attached to
the edge of the pool, the strain rate there being insufficient to produce detachment. In the
second picture from the left, for $\alpha \simeq 70^\circ$, the edge flame is seen to begin to
recede from the pool rim. The last three photographs, from $75^\circ$ to $85^\circ$, document
how the flame eventually begins to lift up out of the boundary layer as the circulation is
increased.

\subsection{Flame lift-off and transition to blue whirl}

After the edge flame lifts up from the pool surface, the presence of the stagnation point near
the bottom of the vortex-breakdown recirculating bubble enables it to stabilize at a new
location, leading to the formation of a lifted partially premixed front belonging to the
blue-whirl family \cite{Xiao.etal.2016, Hariharan.etal.2019a, Hariharan.etal.2019b,
Hu.etal.2019}. Although the shape of the resulting front is drastically different from that
of triple flames propagating along mixing layers, it can be conjectured that the light blue cone
at the base (present even in the far-right frame of the figure but essentially impossible to see
because of excessively weak blue emissions)  corresponds to a rich branch, the bright ring
identifies the stoichiometric point, and the purple haze above is the lean flame, which curves
inwards in response to the surrounding radial velocity. Since the flame is lifted from the
surface, the effect of heat conduction to the fuel is necessarily limited. Hence, with radiation
being also unimportant in these soot-free flames, convection appears to be the only effective
mechanism transferring heat from the flame front to the fuel surface. This observation further
underscores the importance of the flow structure in the collision region at the base of the
flame \cite{Mullen.Maxworthy.1977}.


\section{Conclusions}
\label{sec:conclusions}

It is noteworthy that the relatively simple apparatus described here can produce such a wide
variety of pool-fire and fire-whirl phenomena, merely by varying the single entry-angle
parameter $\alpha$. The experimental observations reported here help to improve our
understanding of fire-whirl phenomena, although it should be abundantly clear from the preceding
discussion that much more research is needed to obtain good quantitative knowledge of these
intriguing phenomena.




\bibliography{references}

\clearpage

\clearpage

\clearpage

\clearpage

\clearpage

\end{document}